\documentclass[a4paper,fleqn]{cas-dc}

\usepackage[numbers,sort]{natbib}
\usepackage{siunitx}
\usepackage{bm}

\newcommand{\titlename}{Competition of magnetic reconnections in self-generated and external magnetic fields}

\begin{document}

\shorttitle{\titlename}
\shortauthors{K. Sakai et al.}

\title[mode = title]{\titlename}

\author[1,2]{K. Sakai}[orcid=0000-0001-9879-9532]
\ead{sakai.kentaro@nifs.ac.jp}

\author[3]{T. Y. Huang}

\author[3,4]{N. Khasanah}

\author[3,5]{N. Bolouki}

\author[3]{H. H. Chu}

\author[1]{T. Moritaka}

\author[6]{Y. Sakawa}

\author[6]{T. Sano}

\author[7]{K. Tomita}

\author[8]{S. Matsukiyo}
 
\author[8]{T. Morita}

\author[2]{H. Takabe}

\author[9]{R. Yamazaki}

\author[1]{R. Yasuhara}

\author[2]{H. Habara}

\author[2]{Y. Kuramitsu}%
\ead{kuramitsu@eei.eng.osaka-u.ac.jp}

\address[1]{National Institute for Fusion Science, 322-6 Oroshicho, Toki, Gifu 509-5292, Japan}
\address[2]{Graduate School of Engineering, Osaka University, 2-1 Yamadaoka, Suita, Osaka 565-0871, Japan}
\address[3]{Department of Physics, National Central University, No. 300, Jhongda Rd., Jhongli, Taoyuan 320, Taiwan}
\address[4]{Universitas Islam Negeri Mataram, Mataram, Indonesia}
\address[5]{Department of Plasma Physics and Technology, Faculty of Science, Masaryk University, Brno 60177, Czech Republic}
\address[6]{Institude of Laser Engineering, Osaka University, 2-6 Yamadaoka, Suita, Osaka 565-0871, Japan}
\address[7]{Graduate School of Engineering, Hokkaido University, Kita 13, Nishi 8, Kita-ku, Sapporo, Hokkaido 060-8628, Japan}
\address[8]{Faculty of Engineering Sciences, Kyushu University, 6-1 Kasuga-Koen, Kasuga, Fukuoka 816-8580, Japan}
\address[9]{Department of Physical Sciences, Aoyama Gakuin University, 5-10-1 Fuchinobe, Sagamihara, Kanagawa 252-5258, Japan}

\begin{abstract}
We investigate the competition of magnetic reconnections in self-generated and external magnetic fields in laser-produced plasmas. The temporal evolution of plasma structures measured with self-emission imaging shows the vertical expansions and horizontal separation of plasma, which can be signatures of reconnection outflows in self-generated and external magnetic fields, respectively. Because the outflows in self-generated magnetic fields are not clear in the presence of the external magnetic field, the external magnetic field can suppress the magnetic reconnection in self-generated magnetic fields.

\end{abstract}


\maketitle

\section{Introduction}

Magnetic reconnection, which rapidly releases magnetic energy into plasmas, triggers various space and astrophysical phenomena such as solar flares, coronal mass ejections, magnetospheric substorms, and auroras \cite{yamada22,shibata11lrsp}. 
It is still unclear how much magnetic energy is converted into bulk flow, thermal, and nonthermal particle energy. The energy partition during a single X-line reconnection has been investigated in space and laboratory plasmas \cite{oka22pop,yamada15pop}. 
In realistic three-dimensional (3D) systems, magnetic reconnections are often highly complex with many X-lines and outflows \cite{daughton11nphys,pyakurel21prl}. For instance, multiple reconnection outflows that interact with particles many times can stochastically accelerate particles to non-thermal energy, which is not taken into account in a single X-line reconnection \cite{hoshino12prl}. The energy partition in multiple X-line reconnection in 3D system can be different from that in a single X-line reconnection and is still an open question \cite{yamada22}. Because there are no precise measurements in space, astrophysical, and laboratory plasmas, we conducted the experiment where multiple magnetic reconnections can take place combining both self-generated and external magnetic fields.
Contrary to numerical simulations, we can experimentally investigate 3D long-time large-scale evolution of magnetic reconnection including kinetic effects, which is one of the most significant advantages of experiments \cite{takabe21hpl}.

\begin{figure}
	\centering
	\includegraphics[clip,width=0.8\hsize]{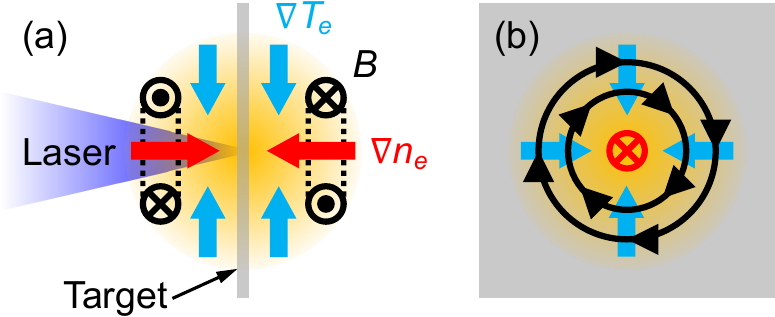}
	\caption{(a) Side view and (b) front view of self-generated Biermann battery magnetic fields.}
	\label{fig:biermann}
\end{figure}

In laser-produced plasmas, magnetic fields are inevitably self-generated due to the Biermann battery effect, which creates an azimuthal magnetic field in the direction of $\nabla T_e \times \nabla n_e$, where $T_e$ is the electron temperature and $n_e$ is the electron density \cite{nilson06prl}. 
When a solid target is irradiated with a laser beam, both temperature and density gradients are generated; the temperature gradient is in the direction toward the focal spot indicated by the blue arrows in Fig. \ref{fig:biermann} due to the localized heating area with the laser beam. The density gradient is in the direction toward the target indicated by the red arrows in Fig. \ref{fig:biermann} because the ablated plasma expands normal to the target surface. Since these gradients are finite and their directions are different, $\nabla T_e \times \nabla n_e$ also has a finite value around the focal spot, generating Biermann magnetic fields.
By separating the focal spots of two beams, one can obtain an anti-parallel magnetic field to investigate a magnetic reconnection. Magnetic reconnections in Biermann magnetic fields have been investigated intensively using two or more laser beams with different focal positions \cite{nilson06prl,zhong10nphys,rosenberg15ncommun,morita22pre,bolanos22ncommun,ping23nphys,khasanah17hedp}. 
On the other hand, magnetic reconnection is investigated by applying external magnetic fields to increase the controllability of magnetic field \cite{fiksel14prl,chien23nphys,zhang23nphys,kuramitsu18ncommun,sakai22srep}. Because one can control the strength of the external magnetic fields, parameters depending on magnetic field strength are well-defined. However, both self-generated and external magnetic fields coexist unless the external magnetic field is so strong that the self-generated magnetic field is negligible. 

We have been working on the electron dynamics in magnetic reconnection \cite{kuramitsu18ncommun,sakai22srep,kuramitsu23rmpp}. In these experiments, a directional laser-produced plasma locally stretches an external magnetic field to form the anti-parallel reconnection configuration. 
We obtained the directional flow on the rear side of a thin foil target by separating the focal spots of laser beams \cite{kuramitsu09apj}. 
Although we use the thin foil target to reduce the Biermann magnetic fields by reducing the temperature gradient at the target \cite{kuramitsu23rmpp}, the separated focal spots can self-generate another anti-parallel magnetic field configuration between the focal spots. 
We performed an experiment on magnetic reconnection in self-generated magnetic fields in the geometry of our electron-scale reconnection experiment \cite{khasanah17hedp}. Although the temporal and spatial scales and magnetic field strength are different from the typical experiments \cite{nilson06prl,zhong10nphys,rosenberg15ncommun,morita22pre,bolanos22ncommun,ping23nphys}, the results show the structure formation in the rear-side plasma due to the magnetic reconnection in the self-generated magnetic fields. Adding an external magnetic field, two magnetic reconnections in self-generated and external magnetic fields can coexist in an experiment. 

In the present paper, we investigate the competition of magnetic reconnections in self-generated and external magnetic fields by comparing results with and without the external magnetic field. We observe plasma structures and physical quantities using optical diagnostics: self-emission imaging, interferometry, and collective Thomson scattering. We observed plasma structures that can relate to the magnetic reconnections in self-generated and external magnetic fields. The magnetic reconnection in self-generated magnetic fields can be suppressed in the presence of the external magnetic field.

\section{Experiment}

\begin{figure}
	\includegraphics[clip,width=\hsize]{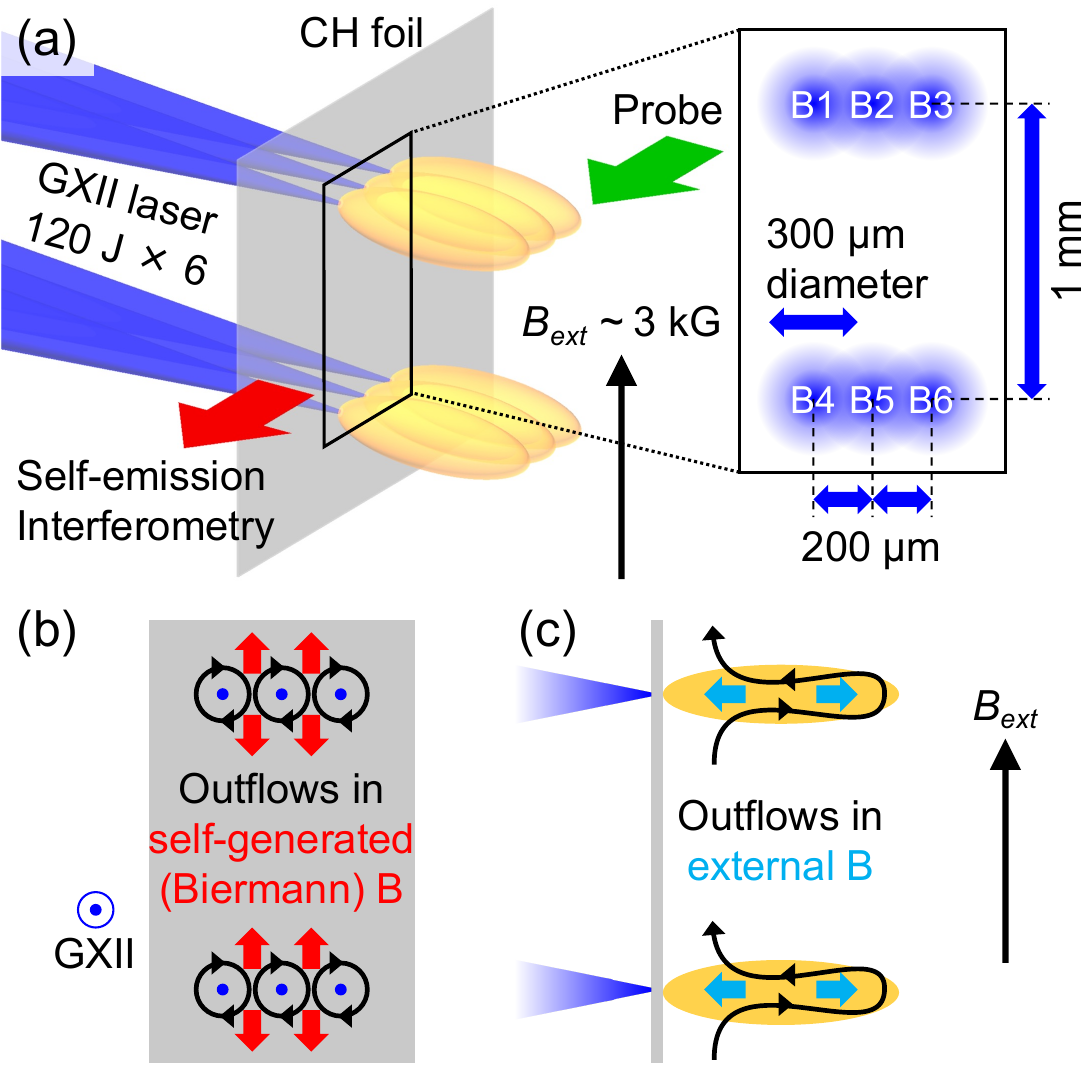}
	\caption{(a) Schematic image of the experimental setup. (b) Front view of the setup. (c) Side view of the setup.}
	\label{fig:setup}
\end{figure}

The experiment was performed with Gekko XII HIPER laser facility at Institute of Laser Engineering, Osaka University. 
Figure~\ref{fig:setup}(a) shows the schematic image of the experiment. A \SI{10}{\micro m}-thick polystyrene (CH) foil is irradiated with 6 main laser beams (B1--B6) with separated focal spots at the same time. The target density is \SI{1}{g/cm^3} and the composition ratio of hydrogen to carbon is unity. The main laser beams have the energy of 120 J per beam, the wavelength of 351 nm, the pulse duration of 500 ps, the focal spot diameter of \SI{300}{\micro m}, and the intensity of \SI{5e14}{W/cm^{2}}. We used the Gaussian spatial and temporal profiles of the laser beams. We define the time when the self-emission begins as the time origin of each detector, which corresponds to the rise of the laser pulse. 
The vertical and horizontal beam separations are 1~mm and \SI{200}{\micro m}, respectively. We made plasma collimation using the beam separations to stretch the external magnetic field locally \cite{kuramitsu09apj,kuramitsu16pop}, resulting in the formation of anti-parallel magnetic fields and the magnetic reconnection \cite{kuramitsu18ncommun,sakai22srep,kuramitsu23rmpp}. The vertical beam separation of 1~mm is the same as the previous experiment in which a long-distance plasma collimation is observed \cite{kuramitsu16pop}. The horizontal beam separation of \SI{200}{\micro m} is the same as the previous experiment in which magnetic reconnection in self-generated magnetic fields is observed \cite{khasanah17hedp}. 
The target chamber is filled with 5 Torr nitrogen gas. A permanent magnet applies an external magnetic field perpendicular to the laser propagation direction. The strength of the external magnetic field is $\sim \SI{3}{kG}$ on the target. 

We observed the plasma structure with transverse optical diagnostics: self-emission imaging and interferometry. In this paper, we focus on the rear-side plasma. The time evolution of self-emission images in a single shot was measured with a framing camera (Ultra Neo, NAC Image Technology Inc.). The framing camera captures 12 frames of images every 5~ns from the laser irradiation timing with the integration time of 5~ns. We put a bandpass filter in front of the framing camera to set the measurement wavelength to 450~nm. 
We injected a probe beam whose wavelength is 532~nm and used the transmitted light for the modified Nomarski-type interferometer \cite{benattar79rsi}. The interferogram was captured with an ICCD camera. The signal is integrated over 250~ps.
The local plasma parameters are measured with collective Thomson scattering (CTS). We use the same system as in Ref.~\cite{bolouki19hedp} to observe the ion feature spectrum.

Figure~\ref{fig:setup}(b) shows the front view of the target. An azimuthal magnetic field is generated around a focal spot due to the Biermann battery process \cite{nilson06prl}. Because the focal spots of six beams are separated, the self-generated magnetic fields interact with each other and the magnetic reconnection can take place between each of the two beams \cite{nilson06prl,zhong10nphys,rosenberg15ncommun,morita22pre,bolanos22ncommun,ping23nphys,khasanah17hedp}. The horizontal beam separation is smaller than the vertical one. The magnetic reconnection in vertical beams can be suppressed since the magnetic energy within the reconnection region is lower than that in horizontal beams. The reconnection outflows are expected to propagate vertically [red arrows in Fig.~\ref{fig:setup}(b)].

In the absence of the external magnetic field, only the magnetic reconnection in self-generated magnetic fields is possible as shown in Fig.~\ref{fig:setup}(b). When we add the external magnetic field, the plasma flow expanding from the target surface can advect and stretch the external magnetic field, as illustrated in Fig.~\ref{fig:setup}(c) \cite{moritaka16pop,vanzeeland04pop}. This results in the formation of an anti-parallel magnetic field and the magnetic reconnection in the external magnetic field can take place \cite{kuramitsu18ncommun,sakai22srep,kuramitsu23rmpp}. The reconnection outflows are expected to propagate along the laser propagation axis [blue arrows in Fig.~\ref{fig:setup}(c)] and the plasma can separate in the horizontal direction.

\section{Results}

\begin{figure*}
	\includegraphics[clip,width=\hsize]{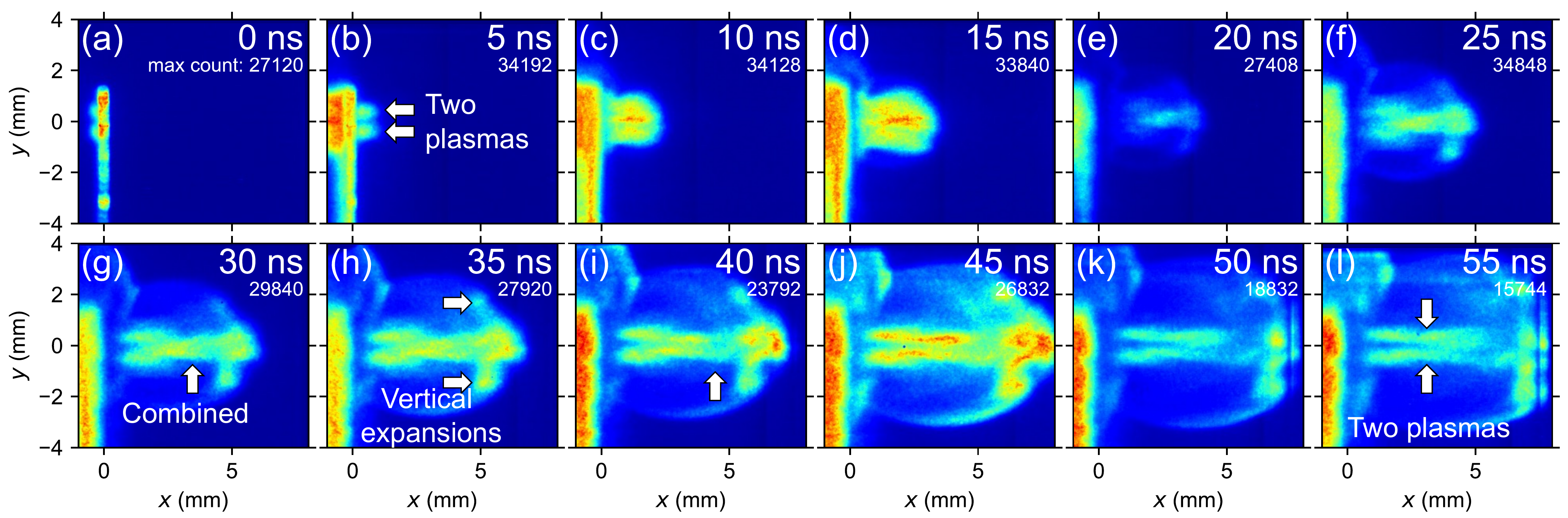}
	\caption{Time evolution of self-emission images in a 6-beam (B1--B6) shot without the external magnetic field. (a)--(l) are the results taken at 0--55~ns, respectively.}
	\label{fig:un_6no}
\end{figure*}

\begin{figure*}
	\includegraphics[clip,width=\hsize]{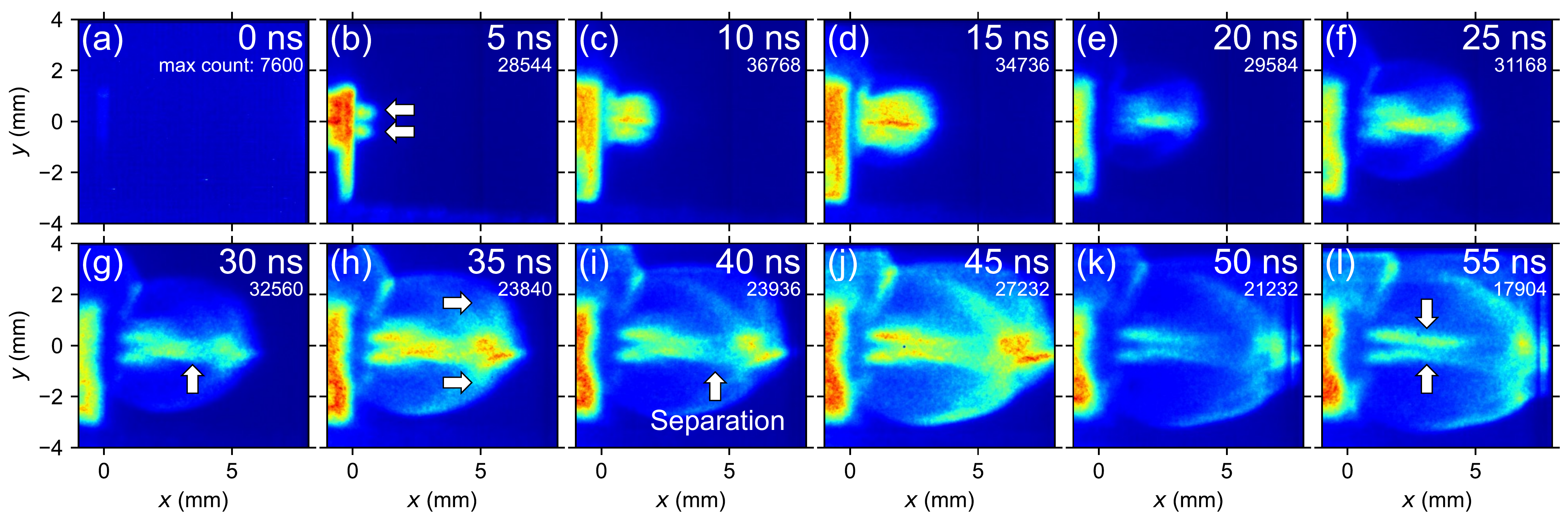}
	\caption{Time evolution of self-emission images in a 6-beam (B1--B6) shot with the external magnetic field. (a)--(l) are the results taken at 0--55~ns, respectively.}
	\label{fig:un_6with}
\end{figure*}

Figure~\ref{fig:un_6no} shows the time evolution of self-emission images measured with the framing camera without the external magnetic field. Figures \ref{fig:un_6no}(a)--\ref{fig:un_6no}(l) are the images taken at 0--55~ns after the main laser irradiation, respectively. The measured timing and the maximum count are shown in the upper right of each image. Figure~\ref{fig:un_6with} is the same plot as Fig.~\ref{fig:un_6no} but with the external magnetic field. The target is located at $x\sim 0$ and the 6 laser beams (B1--B6) propagate from left to right. 
Yellow or red parts of these images correspond to the self-emission in the plasma. Three types of emission can be seen in this image; one is from the ionized target. Since the density near the target is high, the intensity of the emission is the highest in each image. The strong emission propagating along the $x$ axis is the expanding rear-side plasma from the target surface. The elliptically shaped weak emission at the edge of the rear-side plasma is from the ambient nitrogen gas plasma. The target is ionized by the main laser, and the ambient nitrogen is ionized by X-ray emitted at the target surface. There are shocks in the nitrogen plasma. 
The 12 frames of the framing camera have different sensitivities and the color is sometimes weak to see the structure. Because the maximum count approximately corresponds to the emission intensity at the target, we set the maximum color with the maximum count so that the signal intensity is normalized to the intensity at the target. 
The minimum color is set to the dark-current noise which is the background count without any emission signal.
There are two rear-side plasmas near the target at 5~ns, indicated by white arrows in Figs.~\ref{fig:un_6no}(a) and \ref{fig:un_6with}(a). The separation of the two plasmas is consistent with the vertical beam separation of 1~mm [see Fig.~\ref{fig:setup}(a)].
The two plasmas are combined after 10~ns. The combined plasma propagates in the middle of the two plasmas as indicated by the arrow in Figs.~\ref{fig:un_6no}(g) and \ref{fig:un_6with}(g).
After 40~ns, the combined plasma is vertically separated into two plasmas again, which is indicated by the arrows in Figs.~\ref{fig:un_6no}(l) and \ref{fig:un_6with}(l).
The time evolution of the rear-side plasma front is similar regardless of the presence and absence of the external magnetic field. The velocity of the rear-side plasma is $\sim \SI{200}{km/s}$ from the images.
There are vertical expansions behind the front of the rear-side plasma after 25~ns, which are indicated by the arrow in Figs.~\ref{fig:un_6no}(h) and \ref{fig:un_6with}(h). Similar structures are observed in the previous experiment without an external magnetic field \cite{khasanah17hedp}. The emission intensity of the vertical expansions in Fig.~\ref{fig:un_6no}(h) is stronger than that in Fig.~\ref{fig:un_6with}(h). The velocity of the vertical expansions is $\sim 20$--50~km/s from the images. 
The rear-side plasma is separated along the $x$ axis after 40~ns as shown in Figs.~\ref{fig:un_6no}(i) and \ref{fig:un_6with}(i). The horizontal separation occurs at the left of the vertical expansions. 
Since the emission intensity at $x\sim 4$--5~mm in Fig.~\ref{fig:un_6with}(i) is weaker than that in Fig.~\ref{fig:un_6no}(i), the horizontal separation is clearer in the presence of the external magnetic field.
\begin{figure*}
	\includegraphics[clip,width=\hsize]{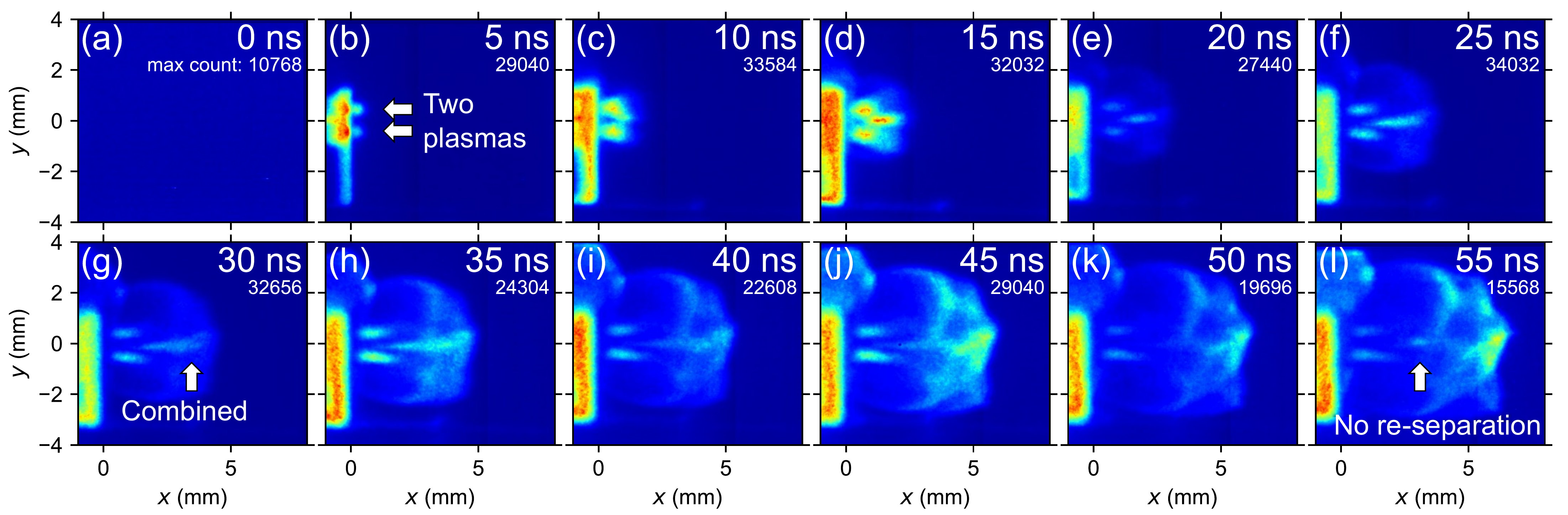}
	\caption{Time evolution of self-emission images in a 2-beam (B2, B5) shot with the external magnetic field. (a)--(l) are the results taken at 0--55~ns, respectively.}
	\label{fig:un_2with}
\end{figure*}
For comparison purposes, we show another result with only two beams divided vertically. Figure~\ref{fig:un_2with} is the same plot as Fig.~\ref{fig:un_6with} except for using B2 and B5 beams. When we irradiate the target with only two laser beams divided vertically, the vertical re-separation indicated by the arrow at 55~ns is not observed as shown in Fig.~\ref{fig:un_2with}(l).

\begin{figure}
	\includegraphics[clip,width=\hsize]{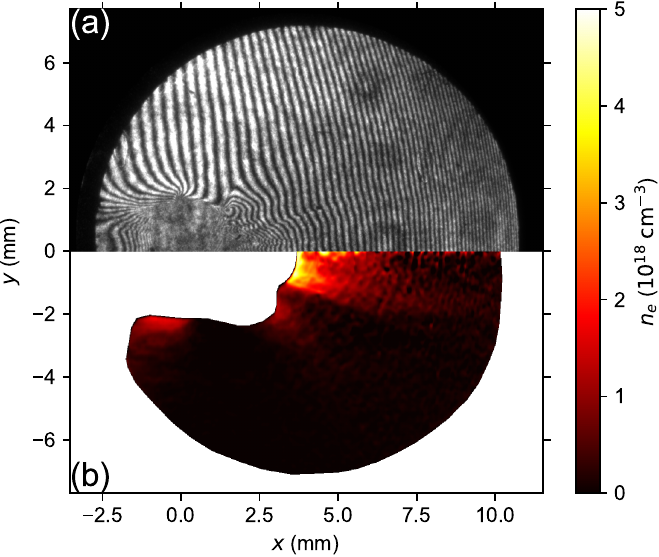}
	\caption{(a) An interferogram with the external magnetic field at 45~ns. (b) Calculated electron density profile.}
	\label{fig:if}
\end{figure}

Figure~\ref{fig:if}(a) shows an interferogram taken at 45~ns with the external magnetic field. The target position is at $x=0$ in this image and the 6 laser beams (B1--B6) propagate from left to right. Note that the image is taken in ``vacuum'' shot ($\sim 10^{-4}$~Torr) to avoid the shocks, or density jumps, in nitrogen plasma. This allows us to calculate the electron density map from the phase shift. 
We assume cylindrical symmetry and perform the Abel inversion to calculate the electron density map in Fig.~\ref{fig:if}(b). The white region in Fig.~\ref{fig:if}(b) is the masked region for calculating phase difference because the fringe is too dense to count. As in Figs.~\ref{fig:un_6no} and \ref{fig:un_6with}, the rear-side plasma is collimated in the $x$ direction. The electron density in the rear-side plasma is $n_e \sim \SI{e18}{cm^{-3}}$.

\begin{figure}
	\includegraphics[clip,width=\hsize]{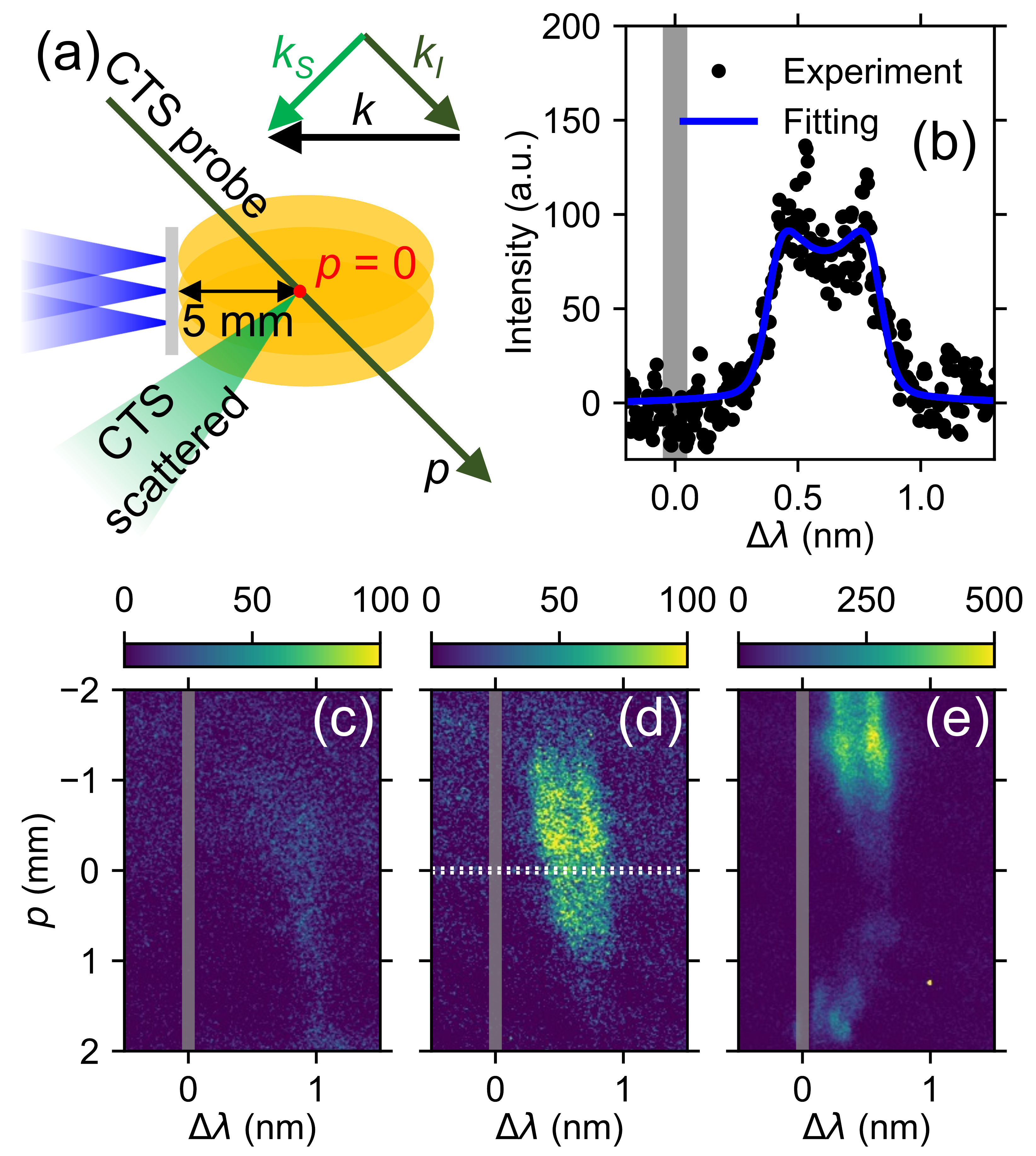}
	\caption{(a) Top view of the setup. (b) Scattered spectrum at $p=0$ with the external magnetic field at 45~ns. Spatially-resolved CTS spectra in (c) 6 beam shot without the external magnetic field at 35~ns, (d) 6 beam shot with the external magnetic field at 45~ns, and (e) 2 beam shot with the external magnetic field at 45~ns.}
	\label{fig:cts}
\end{figure}

Figure~\ref{fig:cts} shows the CTS results. The geometry of the measurement system is schematically illustrated in Fig.~\ref{fig:cts}(a). The wavevector of measured density fluctuation ($\bm{k} = \bm{k_S} - \bm{k_I}$) is parallel to the plasma propagation, where $\bm{k_I}$ and $\bm{k_S}$ are the incident and scattered wavevectors, respectively. We measured the spatial distribution of the scattered spectrum along the CTS probe at 5~mm away from the target on the $y\sim 0$ plane. 
Figures~\ref{fig:cts}(c)--\ref{fig:cts}(e) show the spatially-resolved CTS spectra at the same shots as in Figs.~\ref{fig:un_6no}--\ref{fig:un_2with}, respectively. The measured timing is 35~ns in Fig.~\ref{fig:cts}(c) and 45~ns in Figs.~\ref{fig:cts}(d) and \ref{fig:cts}(e). The horizontal and vertical axes correspond to the wavelength shift and position along the CTS probe, respectively. The wavelength region of the notch filter is shown in gray. The scattered intensity, which is positively correlated with the electron density, is weak in 6 beam shots in the presence of the external magnetic field. The spectrum at $p=0$ in the 6 beam shot with the external magnetic field [white dotted region in Fig.~\ref{fig:cts}(d)] is plotted in Fig.~\ref{fig:cts}(b). The spectrum is fitted with the dynamic structure factor \cite{froula11} assuming the electron density of \SI{1e18}{cm^{-3}} (obtained from interferometry), the same mean velocity of electrons and ions, and the ionization state from the FLYCHK code \cite{chung05hedp}. 
The parameters estimated by nonlinear least squares fitting are as follows; 
the electron temperature is $T_e \sim 20 \pm \SI{2}{eV}$, ion temperature is $T_i \sim 200 \pm \SI{20}{eV}$, the flow velocity is $u \sim 240 \pm \SI{2}{km/s}$, and the ionization state is $Z_p \sim 1$ for proton and $Z_c \sim 4$ for carbon. The uncertainty is estimated by the standard deviation of the fitted parameters.

\section{Discussion}

We estimate the kinetic and magnetic pressures from the interferometry and CTS results. The kinetic pressure is $p_K \sim \sum_i n_i m_i u^2/2 \sim \SI{1.3e9}{erg/cm^{3}}$, where $n_i$ is the ion density and $m_i$ is the ion mass. The magnetic pressure in the external magnetic field is $p_B = B^2/(8\pi) \sim \SI{4e5}{erg/cm^{3}}$, where $B$ is the magnetic field strength. 
Since the magnetic pressure is three orders of magnitude smaller than the kinetic pressure, the plasma motion can advect the external magnetic field resulting in the formation of anti-parallel magnetic field configuration as in Fig. \ref{fig:setup}(c) \cite{moritaka16pop,kuramitsu18ncommun}. The gyroradii of electrons, protons, and carbons are \SI{50}{\micro m}, 1~cm, and 2.5~cm, respectively. Since the system size of our experiment is $\lesssim \SI{1}{cm}$, only the electrons are magnetized in the external magnetic field. 
Because the magnetic flux is frozen into the electrons, the external magnetic field can be transferred and amplified by the electron flow at the plasma expansion front where a shock is formed in the ambient plasma. Moreover, the external magnetic field becomes weak near the target by the diamagnetic cavitation due to the high kinetic pressure of target plasma \cite{moritaka16pop,vanzeeland04pop}.
The gyroradius and magnetic pressure estimated here are the typical values without magnetic amplification and diamagnetic cavitation. Even if the magnetic field changes by one order of magnitude, the electrons are still magnetized and can advect the external magnetic field because the electron gyroradius and the magnetic pressure are $\sim 10^2$ and $\sim 10^3$ times smaller than the system size. In order to understand the effect of magnetic amplification and diamagnetic cavitation due to the plasma flow into magnetic reconnection, it is necessary to measure the magnetic field at the reconnection layer in some way like ion radiography \cite{schaeffer23rmp}.

The typical strength of self-generated magnetic field is $\sim \SI{1}{MG}$ with $\sim \SI{1}{mm}$ spatial scale in early timing ($\sim 1$--10~ns) \cite{nilson06prl,zhong10nphys}. In the present paper, we focus on late time ($\sim 10$--100~ns) and the spatial scale is $\sim \SI{1}{cm}$. When we assume the plasma advects the magnetic field, the mean magnetic field strength is $\sim \SI{1}{kG}$ because of the $10^3$ times larger volume \cite{kuramitsu23rmpp}. Since this is comparable to the strength of the external magnetic field, only the electrons can be magnetized in the self-generated magnetic fields as in the external magnetic field.

The emission of the rear-side plasma is horizontally separated as indicated by the arrow in Fig.~\ref{fig:un_6with}(i). This is similar to the previous experiment \cite{kuramitsu18ncommun}. Therefore, it is considered that the magnetic reconnection in the external magnetic field occurs and the plasma horizontally separates as a result of magnetic reconnection \cite{kuramitsu18ncommun}. The separated point can be the reconnection point. The CTS spectrum in Fig.~\ref{fig:cts}(e) shows a similar spatial distribution as the previous experiment that observed electron outflow without ion outflow in the electron-scale magnetic reconnection \cite{sakai22srep}.

The vertical expansions at $t \gtrsim \SI{25}{ns}$ in Figs.~\ref{fig:un_6no} and \ref{fig:un_6with} propagate in the direction of the reconnection outflows in self-generated magnetic fields as illustrated in Fig.~\ref{fig:setup}(b) \cite{khasanah17hedp}. The timescale of magnetic reconnection is $\sim 25$~ns with the beam separation of \SI{200}{\micro m}, which is consistent with the previous experiment \cite{khasanah17hedp}. This indicates that the magnetic reconnection in self-generated magnetic fields is possible in the presence of the external magnetic field. 
Since the emission intensity in the outflow region without the external magnetic field is stronger than that with the external magnetic field, the magnetic reconnection in the self-generated magnetic fields is suppressed in the presence of the external magnetic field. 
This can be interpreted that a part of magnetic energy in the self-generated fields is released by the magnetic reconnection in the external magnetic field and the magnetic reconnection in self-generated magnetic fields becomes less significant in the presence of the external magnetic field.

In electron-magnetized plasmas, the outflow speed of magnetic reconnection is up to the electron Alfv\'en speed ($c_{Ae}=B/\sqrt{4\pi n_e m_e}$) \cite{hoshino01jgr}. In Fig.~\ref{fig:un_6no}, the velocity of vertical expansions is 20--50 km/s. Assuming the vertical expansion as reconnection outflows at the electron Alfv\'en speed, the strength of the self-generated magnetic field is estimated to be 200--500~G. Because the electron Alfv\'en speed is the maximum outflow speed, the self-generated magnetic field can be even stronger than the estimation and comparable to the external magnetic field strength of 3~kG. Note that the maximum strength of the self-generated magnetic field should be locally stronger than the external magnetic field strength for the magnetic reconnection to occur in self-generated magnetic fields. The self-generated magnetic field is not so weak as to neglect and the competition of self-generated and external magnetic fields is important in the present experiment.

The rear-side plasma is vertically separated in early time ($\sim \SI{5}{ns}$) due to the vertical beam separation. The two plasmas are combined into one collimated plasma at 10--35~ns \cite{kuramitsu09apj}. After 40~ns, the combined plasma is vertically separated into two again in Figs.~\ref{fig:un_6no} and \ref{fig:un_6with} Since the vertical re-separation is not observed in the 2 beam shot (Fig.~\ref{fig:un_2with}), the re-separation is created by the horizontal beam separation. The CTS spectra in the 6 beam shots [Figs.~\ref{fig:cts}(c) and \ref{fig:cts}(d)] show less intensity than that in the 2 beam shot [Fig.~\ref{fig:cts}(e)]. Because the scattered intensity is proportional to the electron density, the electron density is lower in the 6 beam shots in spite of the higher total laser energy. This also suggests that the plasma vertically separates with the horizontal beam separation. In the presence of the external magnetic field, the scattered intensity or the electron density is lower than that without the external magnetic field, so the vertical re-separation is suppressed with the external magnetic field. 
According to the fact that the vertical re-separation is created by the horizontal beam separation, there should be an effective vertical repulsive force due to the horizontal beam separation. One of the candidates to make the horizontal repulsion with vertical beam separation is the magnetic reconnection in the Biermann magnetic fields. In a previous experiment with only two horizontally separated beams, we observed outflows due to magnetic reconnection in the Biermann magnetic fields illustrated in Fig.~\ref{fig:setup}(b) \cite{khasanah17hedp}. Given the results presented here and previously published, it seems most plausible that the reconnection outflows in the Biermann magnetic fields propagating perpendicular to the plasma plumes make an effective horizontal repulsion.

In the presence of the external magnetic field, the magnetic reconnection in self-generated magnetic fields can be asymmetric because the upstream magnetic field strength is asymmetric. 
The asymmetric reconnection potentially modifies the structure of magnetic reconnection, e.g., the position of the X-line is different from that of the stagnation point \cite{cassak07pop}. In order to understand the stagnation point or the velocity structure in the inflow region, we need another Thomson scattering system that measures another component of plasma velocity. Since we are developing the Thomson scattering system in two directions \cite{morita22pre}, it is possible to measure the spatial distribution of the inflow velocity in the electron-scale asymmetric reconnection experimentally. 
Our experiment is at high-$\beta$, which is similar to the experiment in Ref. \cite{rosenberg15ncommun}. In high-$\beta$ asymmetric reconnections, the reconnection rate is determined by the external drivers and not sensitive to the asymmetry \cite{rosenberg15ncommun}, which is consistent with our time-evolution measurements of emission images.

The magnetic reconnection in the self-generated magnetic fields can occur between the upper and lower beams, resulting in the vertical plasma merging and the outflow in the direction toward and against the optical imaging measurements. It is known that the magnetic reconnection in the self-generated magnetic fields does not occur with a large beam separation \cite{willingale10pop}. Because the vertical beam separation is 5 times larger than the horizontal one, the magnetic field strength at the interaction point in vertical beams can be at least one order of magnitude weaker than that in horizontal beams and the external magnetic field strength.

\section{Summary}
We investigated the competition of magnetic reconnections in self-generated and external magnetic fields. 
Without the external magnetic field, the plasma propagates in the vertical direction at $t\gtrsim \SI{25}{ns}$. This is consistent with our previous experiment \cite{khasanah17hedp} and can be reconnection outflows in self-generated magnetic fields. 
On the other hand, with the external magnetic field, the plasma is horizontally separated after 40~ns. This can be a signature of the magnetic reconnection in the external magnetic field at the separated point \cite{kuramitsu18ncommun}. The separation of focal spots affects the long time evolution of laser-produced plasmas due to magnetic reconnection.
When both self-generated and external magnetic field exists, the magnetic reconnection in self-generated magnetic fields can be suppressed. 
Using this platform, we are going to measure and discuss the energy partition during multiple X-line reconnection in 3D.

The competition of external and self-generated magnetic fields can create complex magnetic field lines in 3D, which is composed of, for example, twisted Biermann battery fields and the compressed external magnetic field together with diamagnetic cavitation. 
In order to understand the total magnetic field structure, which is a combination of external and self-generated magnetic fields, we need to measure the 3D distribution of magnetic fields. Ion (proton) radiography is a standard diagnostic to measure electric and magnetic fields \cite{schaeffer23rmp}, however, measuring the 3D electric and magnetic fields with complex structures using the radiograph data is difficult because there is no plausible model to explain the structure. 
We are trying to measure the 3D distribution of electric and magnetic fields with ion radiography. We observe the positions and energies of numerous ions using solid-state nuclear track detectors \cite{hihara21srep} and reconstruct electric and magnetic fields in 3D space with the aid of artificial intelligence \cite{kuramitsu24pop,jao24aadv}.

\section*{Acknowledgements}
The authors would like to acknowledge the dedicated technical support by the ILE staff for the laser operation, target fabrication, and plasma diagnostics. 
This work is supported by JSPS KAKENHI Grant Nos. 24K17029, 23H01162, 23H01211, 22H01195, 22H01251, 21J20499, JSBP120203206, 20KK0064, 19K21865, 15H05751, and Ministry of Science and Technology of Taiwan Grant Nos. 103-2112-M-008-001-MY2, 104-2112-M-008-013-MY3, and 105-2112-M-008-003-MY3. 
The authors appreciate NINS program of Promoting Research by Networking among Institutions (01422301, 01412302) and the grant of Joint Research by NINS (OML022405).
N. B. acknowledges that he was supported by the project LM2018097, funded by the Ministry of Education, Youth and Sports of the Czech Republic. 

\bibliographystyle{model1-num-names}
\bibliography{recomp}

\end{document}